\documentclass[11pt]{article}

\usepackage{amsmath,amssymb,amsthm}
\usepackage{geometry}
\usepackage{hyperref}
\usepackage{cite}
\usepackage{bbm}
\usepackage{mathtools}

\title{\large\bf Why There is No Memory Burden in Holographic Space-time Models of Black Hole Formation and Evaporation}

\author{T.~Banks\\[4pt]
\small NHETC and Department of Physics and Astronomy\\
\small Rutgers University, Piscataway, NJ 08854}

\date{}

\begin{document}
\maketitle

\begin{abstract}
Recent papers\cite{dvali} have claimed that general quantum information considerations put macroscopic constraints on the properties of large black holes, which can affect their lifetimes in cosmologically and astrophysically interesting ways.  We examine black hole evaporation in Holographic Space Time (HST) models\cite{hstbh}, which appear to have the {\it memory burden} effect responsible for these deviations from semi-classical expectations, and explain why, in those models, no such effect exists.  It is basically a consequence of Fermi's Golden Rule/The Principle of Detailed Balance, but depends crucially on both the definition of energy in HST models and the way in which causality is implemented.  
\end{abstract}

\tableofcontents

\section{Introduction}
In a recent series of papers\cite{dvali}, Dvali and collaborators have argued that black holes have a {\it memory burden}, which prevents them from evaporating at exactly the rate predicted by Hawking.  The memory burden effect is supposed to be substantial even for black holes that can have significant effects on astrophysics and cosmology.  The basic intuition is that a black hole can be modeled by a large quantum computer, whose q-bits can be frozen but cannot simply disappear.  For the present author, a clue to understanding the concept of memory burden came from the claim that the open string degrees of freedom in the string theory description of BPS black holes played the role of memory burden, but that one couldn't see the memory burden effect because those black holes are stable.  

HST models of black hole formation and decay\cite{hstbh} were built by analogy with open string descriptions of many black brane systems in string theory.  For readers who have not followed any of the development of these models, we'll provide a summary of the basics in the third section of this paper.  In the second section we'll discuss a ``toy model" which has all the basic ingredients.  Its missing only extra indices that provide information about compact dimensions (which string theory tells us must be present) and the complicated description of the same Minkowski space-time from the point of view of all possible geodesics in it.  The toy model describes the formation and evaporation of a black hole as seen on a geodesic in the black hole rest frame centered at the black hole position.  It also imposes initial conditions at a large finite proper time in the past.  The mathematical subtleties involved in passing to the true conformal boundary of Minkowski space will be outlined in the third section.

\section{Toy Model of Black Hole Formation and Evaporation}

We will work in $4$ dimensional space-time.  The generalization to higher dimensions will be briefly indicated in the next section.  Our toy model stands on its own as a model of memory burden and the fact that it does not affect the lifetime of high entropy meta-stable states in a finite quantum model.  In this section we will content ourselves with simply listing a few parallels with space-time physics.  A more complete picture will be given in subsequent sections.

\begin{itemize}

\item The Dirac fields of our model are in one to one correspondence with a finite set of spinor spherical harmonics on the two sphere.  Following the ideas of Connes\cite{connes} and the HST formalism, combined with those of\cite{CS}\cite{BZ}\cite{tbwflm} these can be thought of as parametrizing fluctuations of the ``fuzzy" geometry of causal diamonds in Minkowski (or de Sitter)\footnote{The difference between Minkowski and dS space-times is implemented in the model by controlling the rate at which the number of fields associated with a causal diamond increases with increasing proper time.}.
\item We'll define density matrices for subsystems of our system of Dirac fields and a sequence of unitary embeddings between the Hilbert spaces of those subsystems, which obey the rules of\cite{CS}\cite{BZ}\cite{tbwflm} relating the central charge of a (cut-off) $ 1 + 1$ conformal field theory (CFT) to the maximal area on causal diamond boundaries. These embeddings are the analogs of half sided modular inclusion in Algebraic QFT (AQFT), which is generated by a sequence of unitaries ({\it not} the global Hamiltonian flow) on the full Hilbert space.  The construction of the corresponding sequence of unitaries is the biggest lacuna in the HST formalism and will be discussed in subsequent sections.  The time evolution implied by these unitaries incorporates the Milne redshift between dynamics on the stretched horizon and that measured by a detector on the geodesic that defines the sequence of diamonds.
\item Scattering states are defined by imposing constraints on the initial state, such that small subsystems do not initially interact.  These states carry an asymptotically conserved quantum number, that is the analog of energy. The dynamics of the model then exhibits analogs of the scattering of localized objects, including change in the number of such objects, as well as black hole (BH) production in scattering, followed by the decay of the ``black holes".  In the model ``BH" simply stands for a high entropy meta-stable equilibrium state, with energy, entropy and lifetime that are related in the same way as those of a black hole in $4$ dimensional Minkowski space.  

\item The constraints on initial states, for entropy deficit much smaller than the total entropy of the system, give an entropy deficit like that of the Schwarzschild de Sitter black hole.  Following these constraints through smaller subsystems determines the difference between ``particle scattering amplitudes" and ``black hole production in particle scattering".  The constraints can be given a geometrical interpretation, which will be explained in the next section.

\end{itemize}

 The complete set of degrees of freedom of our system consists of $M(M + 1)$ Dirac fermion fields, $\psi_i^A$,   living on the interval $[0,\pi]$ with reflecting boundary conditions.  The momentum spectrum is cut off at about $20$ modes, the place where Cardy's asymptotic spectral density formula begins to be a good approximation. 
 
In the {\it empty diamond state}, the analog of the QFT vacuum, we postulate that the density matrix in each diamond is 
\begin{equation} \rho^{N} = e^{ - L_0 (N) + \frac{g^2}{N^2} {\rm Tr} \int dz J^m (z) J_m (z)}, \end{equation} where 
\begin{equation} (J)_j^{i\ m} = \bar{\psi}^i_A \gamma^m \psi_j^A , \end{equation} is the $U(N_{\pm})$ current formed from the fermions.  If we view the fermions as eigenspinors of the Dirac operator on the horizon two sphere, then fermion bilinears are differential forms, and the $U(N)$ symmetry acts like a ``fuzzy" approximation to the transformations of area preserving maps of the two sphere.  

Consider some subset of the fermion fields, with labels in some basis running from $1$ to $N$.  We'd like to define a unitary embedding of the Hilbert space of these variables into the larger space with indices up to $N + 1$, which maps the empty diamond density matrix $\rho (N)$ into $\rho (N + 1)$.  These are the analogs of {\it half sided modular inclusion} in Algebraic Quantum Field Theory.   Although in the near horizon coordinate frame, this evolution must take place in one Planck time, the Milne red shift tells us that the geodesic proper time for evolution is $1/N L_P$.  So the time dependent Hamiltonian for geodesic time evolution is rescaled by a factor of $(1/N) M_P$ relative to the generator of half sided modular flow that maps $\rho_{\diamond} (N)$ into $\rho_{\diamond} (N + 1)$ in a time of order $L_P$.   In AQFT, half sided modular inclusion in a set of nested causal diamonds is generated by a sequence of unitary operators acting on the whole Hilbert space.  

In QFT the Hilbert space does not factorize into a tensor product of operators localized in a diamond and those outside it.  In our finite dimensional model there is such a factorization. Fermion fields $\psi_i^A$ with $i > N + 1$ or $A > N + 2$ anti-commute with all of the fermions with indices in the lower range.  So we can insist on a version of sharp causality, namely that the unitary evolution operator on the full Hilbert space over the time interval $[- T + N L_P , - T + (N + 1) L_P]$, as seen by a detector on the geodesic, is the tensor product of the unitary embedding of the Hilbert space of fermions with indices $(N,N+1)$ into that with indices $(N+1,N+2)$ with an operator $U_{out} (N)$ that acts on the rest of the fermions.  It's natural to associate $\psi_i^A$ with {\it all} indices greater than $(N,N+1)$ with the region shown in Figure 1.  This is not a causal diamond because its past null boundary does not converge to a single point.  Thus the question of what how the evolution operator $U_{out} (N)$ should be chosen is not completely clear.  We'll see that, for the purposes of the present section, this is irrelevant, as long as that choice is consistent with certain assumptions about the boundary conditions on subsystems as a function of time $N$.

To define a scattering theory using these notions of time evolution inside nested subsystems, we need two rules.  The first is that we start at time $- T$, and evolve forward in time with the {\it inverse} of the embedding maps defined above, until $t = 0$, and then proceed to $t = T$ by performing the sequence of embedding maps in their proper order.  Since embedding maps are not invertible, for a given set of initial conditions, there are many ways that the system could evolve, and those depend on the undetermined $U_{out} (N)$.  In the next section, we'll see that by viewing this model as evolution along a geodesic in Minkowski space-time, we can in principle determine $U_{out} (N)$.  Here we'll instead define ``energy" and ``scattering states" in terms of constraints on the initial state of the fields $\psi_I^A$,and make assumptions about how the time evolution evolves the constraints onto smaller and smaller subsystems.

Since the empty diamond modular Hamiltonians are all single trace operators, it is consistent to assume that the time dependent Hamiltonian operator that performs the modular embeddings is also single trace.  We define scattering states at time $t = - T$ by insisting that, when applied to those states, all bilinear operators of the form
\begin{equation} \bar{\psi}_i^A (x) \Gamma \psi_A^j (y) , \end{equation} are block diagonal, with a finite number of blocks of size $m_i $ such that $\sum m_i \ll T/L_P$ ,  and one large block.  It then follows that the quantity $\sum m_i$ is an approximately conserved quantum number, which can change only by amounts of order $1/T$.  

This statement is a consequence of two features of our time evolution operator and the constraints.  The constraints require of order $\sum m_i T/L_P$ q-bits to be frozen.  The Hamiltonian that acts on all the degrees of freedom has a natural time scale of $T$.  Hamiltonians that act on smaller numbers of degrees of freedom evolve their subsystem much more rapidly, but cannot change the state of a large number of q-bits, of order $T$. As $T/L_P\rightarrow\infty$, $\sum m_i$ becomes an {\it asymptotically conserved quantum number}.  Below we will identify it as proportional to the energy in Planck units.

The fields $\psi_i^A$ with indices out of the range between $(1,2)$ and $(N,N+1)$ represent, in the space-time picture degrees of freedom in the shaded region shown in Figure 2.  This is not a causal diamond and it is not immediately obvious how to generalize our ansatz for the density matrix and time evolution operator, in order to describe this region.  In the next section we'll see how embedding the system in a Hilbert bundle of systems propagating along other geodesics resolves the problem.

For the purposes of this section though, it is sufficient to formulate the following set of alternative questions about initial value problems.  Given an initial condition at $ - T$ with a set of small block diagonals of sizes summing up to $n$, we can follow the evolution until the total size of the system is $(N,N+1)$ with $ 1 \ll N \ll T/L_P $.  We now ask whether the subsystem bilinears are block diagonal, and what the block sizes are.   There are roughly three possibilities
\begin{itemize}
\item There is no block diagonalization at all.  Since we've argued that a number of constraints of order $n T/L_P$ has to survive the time evolution, those constraints must involve fields not contained in our $(N,N+1)$ subsystem.  We'll see later that the space-time interpretation of this subsystem is a causal diamond along a particular geodesic and this is the statement that none of the initial energy enters that causal diamond.

\item $m$ the sum of the block sizes of small diagonal blocks is $\ll N$.  Then arguments identical to those given above show that the final state of this subsystem again has a small number of small blocks with total size $m$ up to corrections of order $1/N$.  The space-time interpretation of such an interaction is an effective field theory interaction of localized objects, which conserves energy except for emission of ``soft quanta of energy $1/N$".  

\item Finally, if $m \sim N$ we can no longer treat the constrained initial state as a small deviation from the equilibrium empty diamond density matrix.  Since the dynamics was designed to flow to equilibrium, we can expect that such an initial state will quickly flow to the equilibrium empty diamond state for this subset of degrees of freedom.  However, it is important to note that {\it this is far from the empty diamond state of the whole system}.  Indeed, the way we keep track of the integer $m$ is by asking how much of the approximately conserved quantity $n$ of the entire system, has flowed into this particular subsystem given the assumed initial conditions.  This implies that among the degrees of freedom {\it not} belonging to the $(N,N+1)$ subsystem, of order $m (T/L_P)$ q-bits have been frozen in order to create this state.  The state is thus a high energy, high entropy equilibrium state of the system, which we will see in a moment, is meta-stable. 
From the space-time point of view, events of this type correspond to ``black hole creation in particle scattering".

\item  Now let's follow the system to larger time intervals.  For simplicity, assume that $m = n \sim N$ so that all the energy in the initial state went into forming on high entropy equilibrium state at a time $- T + 2 N L_P$.  The fermion bilinear matrices have two blocks of size $N$ and $(T/L_P) - N \gg N$, with off diagonal elements zero, at this time. The off diagonal elements are turned on only a few at a time, at a slower and slower rate as the system size grows\footnote{We're assuming that $U_{out} (N)$ for $N \ll T/L_P$ evolves most operators at a rate $1/T$.}.  The dynamics inside the block of size $N$ has a natural time scale of order $N$.  There is a probability of order $e^{- c pN}$ in the equilibrium density matrix, with $c \sim 1$, to find a bilinear matrices block diagonalized with blocks of size $p$ and $N - p$ .   Once the system has grown to encompass many more fermion fields, there are many more frozen q-bits, since $N$ is the approximately conserved energy.  
\end{itemize}
According to Fermi's Golden rule, we should now be looking at the full space of available final states of energy $N$, in this much larger Hilbert space, where the fermion fields are labeled by $(K, K+1)$, with $K \gg N$. Fermion bilinears are constrained to have of order $NK$ vanishing matrix elements.  The meta-stable bound states have all $N$ of the original fermion fields in one block and there are $K - N$ ways of doing this.  On the other hand if we break $N$ up as $ N = \sum p_i $ with $r$ terms in the sum, we get an extra factor of $\frac{K!}{\prod p_i! (K - r)!} $ from permuting the smaller blocks.  For $K \gg N$ the phase space factor in the probability heavily favors the decay.  Thus, although there's a ``memory burden" in freezing the q-bits in the original meta-stable bound state it's more than made up for by the additional phase space that opens up as modular time evolves.  The fact that our time evolution is a unitary embedding, first run backwards to form the quasi-stable equilibrium state, and then run forwards to watch it decay, is the crucial mechanism that makes all of this consistent with quantum mechanics. 

From the point of view of just modeling the phenomenon of memory burden and showing that it does not affect the decay lifetime predicted by simple calculations of decays into individual channels, we are done.  Any completion of our ``inner" time evolution with a $U_{out} (K)$ acting on the tensor complement of the $(K,K+1)$ Hilbert space, which gives rise to the required initial conditions for formation of the $(N,N+1)$ equilibrium state, is a good quantum model.  In the next section, we'll show that in fact, we can make choices that give us a plausible candidate for real models of quantum gravity in Minkowski space-time.  

\section{A Hilbert Bundle Over Minkowski Space-time}

Given a compactification of Superstring/M-theory on a space-time of the form $M^d \times {\cal K}$, where ${\cal K}$ is compact, the low energy effective SUGRA theory allows us to define a local algebra of operators over the momentum null cone $p^2 = 0$\cite{ags}\footnote{The seminal paper\cite{ags} studied only pure SUGRA.  Superstring/M theory has taught us that below $10$ dimensions SUGRA is always accompanied by a spectrum of stable massive BPS states, as well as some BPS-anti-BPS states predicted by K-theory and its 11 dimensional analog.}
\begin{equation} [\bar{Q}_a^{\pm\ i} (p), Q_b^{\pm\ j} (q)]_+  = \delta (p\cdot q) (\gamma^m )_{ab} p_m \delta^{ij} . \end{equation} 
\begin{equation} [\bar{Q}_a^{\pm\ i} (p), Q_b^{\pm\ j} (q)]_+  = \delta (p\cdot q) (\gamma^m )_{ab} p_m \delta^{ij} . \end{equation} 
\begin{equation} [\bar{Q}_a^{\pm\ i} (p), Q_b^{\mp\ j} (q)]_+  = \delta (p\cdot q) M^{ij} \delta_{ab} . \end{equation} 
\begin{equation} p_m (\gamma^m )_{ab} Q_b^{+\ i} = \tilde{p}_m (\gamma^m )_{ab} Q_b^{-\ i} = 0 .\end{equation} \begin{equation}  \tilde{p} = (p_0, - \vec{p}) . \end{equation}
Since the null cone has a singularity at $p = 0$, there are actually two copies of this algebra, one in the past (which we conventionally identify with negative $p_0$) and the other in the future.  Scattering amplitudes are correlation functions of the operators in this algebra satisfying a number of rules.  Formally they are defined by a state on the pair of algebras, which is the analog of a locally normal state defined in Algebraic Quantum Field Theory to define scattering amplitudes for Nambu-Goldstone bosons and photons.  We will not have to delve into these mathematical complexities.  

The rules for the correlation functions are motivated by the fact that the coefficient of $\delta (p\cdot q)$ in the anti-commutation relations, looks like the relation for super-particles described in terms of one or two (for massive particles) null momenta.  They are Poincare invariance, plus a rule that for non-zero $p$, all the $Q_a^{\pm\ i} (p)$ should vanish outside of a finite number of spherical caps with non-overlapping opening angles.  Each cap should be surrounded by an annulus where $Q_a^{\pm\ i} (0)$ vanishes.  Intuitively, these amplitudes describe scattering of any finite number of finite momentum particles into, with emission of arbitrary amounts of zero energy supergravitons in the initial and final states.  

The indices $i$ represent a basis of eigenspinors of the Dirac operator on the compact manifold ${\cal K}$.  If we believe in the covariant entropy bound\cite{fsb} then there must be an ultraviolet cutoff on Dirac eigenvalues if the volume of the compact manifold is finite in higher dimensional Planck units.  This cutoff is not visible in string perturbation theory but is the source of the {\it stringy exclusion principle}\cite{stringexc} in AdS/CFT.  Correspondingly, the way to describe finite causal diamonds is to restrict the number of Dirac eigenvalues on $S^{d-2}$.  This apparently still leaves the continuous positive parameter $p$ describing the cone over this sphere.

The key to understanding the origin of $p$ in the infinite diamond limit is the relation between the constraints on the operator algebra described above, and the matrix model constraints of the previous section.  The counting of eigenspinors for the Dirac operator on $S^{d-2}$ below some cut-off is the same as the counting of totally anti-symmetric tensors $\psi_{i(1)\ldots i(d-2)}$.  The rectangular matrix of the previous section is just the matrix of independent elements of these for the case $d = 4$.  We can make ordinary square matrices from these by multiplying fields by their conjugates and contracting together $d - 3$ of the indices.  The trace on products of these bilinears is the fuzzy analog of the integral of differential forms.  

If we take the variables describing a causal diamond of size $N L_P$ in $4$ dimensional Planck units as spinor harmonics in the above algebra up to $N$, then a constraint on states that sets $\psi_{i(1) \ldots i(d-3) J} = 0 $ (other indices suppressed for clarity) where small latin indices refer to the first $N$ harmonics and $J$ is all the higher ones up to a cutoff $N*$ that's eventually taken to infinity, sets the off diagonal $U(N*)$ symmetry currents that don't commute with $U(N)$ to zero.  The only interaction, at leading order in time, between the variables in the small diamond and those in larger diamonds comes through the current current interaction between $U(N)$ currents, as the degrees of freedom in larger diamonds are slowly added to the smaller diamond as $N$ increases.  We will describe a localized object in a large causal diamond of size $N*$ in Planck units, as a subset of the degrees of freedom, corresponding to a smaller diamond of size $N \ll N*$, subject to the above constraints.  If we consider the time evolution in the large causal diamond, the constraints and the form of the time evolution operator guarantee that this system will evolve approximately as an independent system for times of order $N*$, if we use the rule that time evolution in a subsystem of size $M$ has a natural time scale of order $M$ (the generalization of 't Hooft scaling to tensor models makes it easy to implement this rule), and the causality preserving rules using time dependent evolution operators (modular inclusions) that factorize between subsets of degrees of freedom.  One then finds, as in the previous section, that for independent localized objects of size $N_i$, the quantity $ \sum N_i^{d-3} $ is an approximately conserved quantity.   We can now consider the limit $N* \rightarrow \infty$ and take $N_i \rightarrow \infty$ with 
$\sum N_i^{d-3} \ll N^{d-3} $ and $\frac{N_i }{ N_j}$ fixed.  This limit introduces a continuous positive variable, which gives us a cone over the sphere, defined in terms of constraints.  Most of the variables carry zero values of this new quantum number, which becomes exactly conserved in the limit.  The limit also restores the $S^{d-2}$ geometry, at least as a measure space, and the ability to have independent localized objects with different values of energy allows us to interpret the localization in terms of angles.

This description pre-supposes that there is a consistent way to factorize the dynamics in this causal manner.  If we're just inventing quantum mechanical models, as we did in the previous section, this is not an issue.  Here we want to make the model consistent with a space-time picture.  The AGS algebras are Poincare covariant, but once we start talking about the algebras of nested finite area diamonds, we are necessarily restricted to the diamonds along a particular time-like geodesic.  The way to re-introduce covariance is to define the quantum theory in a Hilbert bundle over the space of all time-like geodesics.  The natural connection on this Hilbert bundle is furnished by the Quantum Principle of Relativity (QPR): every pair of causal diamonds might have a non-empty intersection and that intersection has a maximal area causal diamond in its interior.  The two diamonds can be along the same geodesic, or two different geodesics.  The QPR says that the intersection diamond is isomorphic to a subsystem of each of the two individual diamonds, with the Hilbert space assigned to its area by the Carlip Solodukhin ansatz.  Furthermore, the time evolution along all geodesics must be such that the entanglement spectra of all of these intersection diamond density matrices must be the same, independently of which geodesic is used to compute the time evolution, for all choices of initial state.  

By construction, the QPR is satisfied for the empty diamond state.  Scattering states are defined by constraints on the empty diamond state.  In a coarse grained way, we can see how the QPR builds up the space-time picture of scattering.  The initial state in some causal diamond of size $N$ has constraints on a small number of the diamond's degrees of freedom, which are invariant under a large invariance group, that approximates the group of volume preserving maps on $S^{d-2}$ .   In order for that state to be consistent, the final state in some set of disjoint causal diamonds just outside that diamond, along geodesics at various angles, must have had a corresponding set of constraints.  This partially fixes the state and the evolution operator $U_{out} (N)$ for the diamond in question.  Similarly, the out states of this diamond determine constraints on the in states of disjoint diamonds in its future.  Extrapolating to the past and future one can see how, at least in a coarse grained way, the QPR completely determines time evolution along all geodesics, and guarantees Poincare invariance.  This argument does not guarantee that the microscopic details work out.  Indeed, the evidence from perturbative string theory and AdS/CFT is that consistent models of quantum gravity in Minkowski space can all be viewed as supersymmetric compactifications of 10 or 11 dimensional SUGRA, and the argument presented here, while it guarantees SUSY, does not see the need for higher dimensions than $4$.  

\section{Conclusions}

We have exhibited models in which the mechanisms that are supposed to be operative in ``memory burden" constraints on black hole evaporation are present, but do not change naive predictions for the evaporation rate.  The meta-stable equilibrium states of these systems indeed decay by ``accidentally" freezing some of their internal q-bits, but the entropy lost in that freezing process is made up for by the large phase space gained by the fragments.  Two unusual features of the dynamics of these models made this possible.  The first was that time evolution was time dependent, mimicking half sided modular flow in quantum field theory, and the second was that all independently evolving subsystems are constrained states of a larger system, with a dynamics that tends towards an equilibrium that removes the constraints.  In the second part of the paper we argued that this kind of dynamics was a plausible model for real theories of quantum gravity, making contact with string theory.  This material has been discussed extensively elsewhere, so the presentation was abbreviated.  

Although our detailed argument depended on specific models, which have not been shown to give a Poincare covariant description of quantum gravity in Minkowski space, they do show that the general idea behind memory burden, that the finite number of q-bits of a quantum computer somehow impede its ability to decay, is not universally valid.  We remind the reader that, in talks, Dvali has often mentioned the open strings between branes in string theory models of BPS black holes as an example of memory burden.  Our off diagonal matrix degrees of freedom are precise analogs of those.  The explicit model of black hole entropy used in most of the papers in\cite{dvali} is based on bulk quantum field theory.  The entropy comes from ``a high-density N-graviton condensate near a quantum phase transition point, where the mass of Goldstone-like modes vanishes, causing quantum backreaction to dynamically slow down state transitions".  There have however been arguments going back at least as far as\cite{ckn}\cite{thooft}\cite{belgiorno}, and more recently strengthened by the {\it firewall paradox}\cite{firewall}, that bulk QFT cannot account for black hole entropy.  Instead, it is neatly explained by the ansatz of\cite{CS}, which we used in this paper.

\end{document}